\documentclass[conference]{IEEEtran}

\makeatletter
\newcommand{\linebreakand}{%
  \end{@IEEEauthorhalign}
  \mbox{}\hfill\par
  \mbox{}\hfill\begin{@IEEEauthorhalign}
}
\makeatother

\usepackage{cite}
\usepackage{algorithmic}
\usepackage{textcomp}
\usepackage{xcolor}
\def\BibTeX{{\rm B\kern-.05em{\sc i\kern-.025em b}\kern-.08em
    T\kern-.1667em\lower.7ex\hbox{E}\kern-.125emX}}

\usepackage{graphicx}
\graphicspath{ {./figures/} }
\usepackage{amssymb}
\usepackage{amsmath}
\usepackage{amsfonts}
\usepackage{hyperref}
\usepackage{multirow}
\usepackage{makecell}
\usepackage{booktabs}
\usepackage{siunitx}
\usepackage{array}

\usepackage{color}

\usepackage{tabularx}
\usepackage{arydshln}
\usepackage{pdflscape}
\usepackage{adjustbox}
\usepackage{float}

\usepackage[T1]{fontenc}

\usepackage{caption}
\usepackage{subcaption}

\usepackage{tikz}
\usetikzlibrary{arrows,shapes,positioning}
\usetikzlibrary{calc,decorations.markings}
\input{macros-generic}
\usepackage{xspace} 

\newcommand{\casePeriodStart}{January, 2021\xspace}
\newcommand{\casePeriodEnd}{July, 2023\xspace}

\newcommand{\fraudcasePeriodStart}{January, 2018\xspace}
\newcommand{\fraudcasePeriodEnd}{August, 2022\xspace}

\newcommand{\sexcasePeriodStart}{January, 2019\xspace}
\newcommand{\sexcasePeriodEnd}{July, 2023\xspace}

\newcommand{\fraudNrClustersWithCollectors}{23\xspace}
\newcommand{\sextortionNrClustersWithCollectors}{81\xspace}
\newcommand{\fraudNrSizeOneClusters}{20\xspace}
\newcommand{\sextortionNrSizeOneClusters}{35\xspace}

\newcommand{\pctLinkedCasesSextortion}{96.9\%} 
\newcommand{\pctLinkedCasesFraud}{41\%} 

\newcommand{\totalnrCases}{1827\xspace}
\newcommand{\totalnrAddresses}{911\xspace}
\newcommand{\sextortionNrCases}{1793\xspace}
\newcommand{\sextortionNrAddresses}{828\xspace}
\newcommand{\fraudNrCases}{34\xspace}
\newcommand{\fraudNrAddresses}{83\xspace}

\newcommand{\totalnrAddressesNoActivity}{554\xspace}

\newcommand{\clustersRemovedToBigOrExchange}{13\xspace}

\newcommand{\totalCasesWithOnchainActivity}{1184\xspace}
\newcommand{\sextortionCasesWithOnchainActivity}{1150\xspace}
\newcommand{\fraudCasesWithOnchainActivity}{34\xspace}
\newcommand{\sextortionAddressesWithOnchainActivity}{275\xspace}
\newcommand{\fraudAddressesWithOnchainActivity}{83\xspace}
\newcommand{\totalNrCaseClustersAddress}{242\xspace}
\newcommand{\totalLargestCaseClusterAddress}{80\xspace}
\newcommand{\fraudNrCaseClustersAddress}{34\xspace}

\newcommand{\sextortionNrCaseClustersAddress}{208\xspace}

\newcommand{\totalNrCaseClustersEntity}{148\xspace}
\newcommand{\totalLargestCaseClusterEntity}{662\xspace}
\newcommand{\fraudNrCaseClustersEntity}{33\xspace}

\newcommand{\sextortionNrCaseClustersEntity}{115\xspace}

\newcommand{\nrvictimsaddressessextortion}{2359\xspace}
\newcommand{\nrvictimsaddressesfraud}{2431\xspace}

\newcommand{\nrtxsextortionextended}{2599\xspace}

\newcommand{\nrtxfraudextended}{4909\xspace}

\newcommand{\nrExpandedSextortionEntities}{147\xspace}
\newcommand{\nrExpandedFraudEntities}{69\xspace}

\newcommand{\nrExpandedSextortionAddresses}{968\xspace}
\newcommand{\nrExpandedFraudAddresses}{1563\xspace}
\newcommand{\nrExpandedAddresses}{2531\xspace}

\begin{document}

\title{Increasing the Efficiency of Cryptoasset Investigations by Connecting the Cases}

\author{\IEEEauthorblockN{Bernhard Haslhofer}
   \IEEEauthorblockA{Complexity Science Hub\\
   Vienna, Austria\\
   email: \textit{haslhofer@csh.ac.at}}
   \and
   \IEEEauthorblockN{Christiane Hanslbauer}
   \IEEEauthorblockA{Magistrate Court of Bamberg\\
   Bamberg, Germany\\
   email: \textit{christiane.hanslbauer@ag-ba.bayern.de}}
   \and
   \IEEEauthorblockN{Michael Fr\"{o}wis}
   \IEEEauthorblockA{Iknaio Cryptoasset Analytics GmbH\\
   Vienna, Austria\\
   email: \textit{michael@ikna.io}}
   \linebreakand
   \IEEEauthorblockN{Thomas Goger}
   \IEEEauthorblockA{Bavarian Central Office for the Prosecution of Cybercrime (ZCB)\\
   Bamberg, Germany\\
   email: \textit{thomas.goger@gensta-ba.bayern.de}}
}

\maketitle


Law enforcement agencies are confronted with a rapidly growing number of cryptoasset-related cases, often redundantly investigating the same cases without mutual knowledge or shared insights.
In this paper, we explore the hypothesis that recognizing and acting upon connections between these cases can significantly streamline investigative processes.
Through an analysis of a dataset comprising \fraudNrCases{} cyberfraud and \sextortionNrCases{} sextortion spam cases, we discovered that \pctLinkedCasesFraud{} of the cyberfraud and \pctLinkedCasesSextortion{} of the sextortion spam incidents can be interconnected. We introduce a straightforward yet effective tool, which is integrated into a broader cryptoasset forensics workflow and allows investigators to highlight and share case connections.
Our research unequivocally demonstrates that recognizing case connections can lead to remarkable efficiencies, especially when extended across crime areas, international borders, and jurisdictions.



\section{Introduction}
\label{sec:introduction}

The increasing prevalence of cryptoasset-related crimes has resulted in a surge of cases that criminal investigators are struggling to manage. At the Bavarian Central Office for the Prosecution of Cybercrime (ZCB), for example, five full-time prosecutors are currently investigating more than 1600 fraudulent cybertrading platforms, with thousands of individual victim cases awaiting attention. These investigations have resulted in victim losses amounting to 250 million Euros in 2022\footnote{\url{https://www.justiz.bayern.de/presse-und-medien/pressemitteilungen/archiv/2022/72.php}} and are estimated to be in the billions in Germany alone.

In addition to cybertrading fraud, investigators are faced with the challenge of addressing cryptoasset abuse, which has become a pervasive issue cutting across various forms of criminal activity. For instance, in ransomware attacks, victims are forced to pay ransom in the hope of receiving the keys necessary to unlock their data, which has been encrypted by the perpetrator~\cite{Paquet:2019a}. Similarly, victims of sextortion spam~\cite{Paquet:2019b} often pay using Bitcoins to prevent the disclosure of potentially damaging adult material, which has supposedly been collected by the perpetrator.

The technique of ``following the money'' is a widely recognized method in forensic investigations, and it has now become a standard approach in cases involving cryptoassets as well. In response to this, a growing industry with a market value of billions of USD has emerged to offer blockchain forensics tools to investigators worldwide. Despite this progress, the current state of affairs remains less than satisfactory: these tools are often prohibitively expensive, and their availability is limited, given the high number of cases that investigators are dealing with. Consequently, numerous cases continue to pile up on the desks of a few investigators, leading to long waiting periods for results. Unfortunately, these waiting periods are frequently too extended, and in many cases, the funds have already been laundered through sophisticated networks that are challenging or impossible to trace manually.

In light of the current situation, it is becoming increasingly evident that merely augmenting manpower and expanding tool licenses is insufficient in addressing a rapidly growing problem. Instead, a more sustainable approach necessitates a period of reflection, a step back, to reevaluate and reimagine how investigations could be conducted more efficiently. One promising avenue stems from the observation that investigators frequently work on identical cases, often without the awareness of each other's efforts or the opportunity to share findings.

In this study, we leverage this observation, hypothesizing that cryptoasset investigations can be organized more efficiently if connections between cases are known. To validate this hypothesis, we undertake an empirical examination using real case data amassed by the ZCB, focusing on cybertrading fraud and sextortion spam. The key contributions of our research are as follows:

\begin{enumerate}

	\item We furnish a dataset consisting of \fraudNrCases{} cybertrading fraud and \sextortionNrCases{} sextortion spam cases with associated cryptoasset addresses.

	\item We propose a straightforward method for identifying connections between cases, premised on the reuse of cryptoasset addresses and the utilization of common collector wallets.

	\item We offer empirical evidence highlighting the prevalent use of cryptoasset in cybercrime. Our findings demonstrate that we automatically find links between \emph{\pctLinkedCasesFraud{}} of the cybertrading fraud and \emph{\pctLinkedCasesSextortion{}} of the sextortion spam cases.

	\item We present a cryptoasset case management solution allowing investigators to collaboratively identify and share connections between cases as part of their forensics investigations.

\end{enumerate}

Our findings corroborate our hypothesis and demonstrate enormous potential for efficiency improvements. Extrapolating from this modest sample of cases originating from a single institution, we assert that substantial efficiency enhancements are attainable if institutions foster collaboration and share fundamental case-related data with each other.

We advocate for a paradigm shift: given the interconnected, global nature of cybercrime, our approach should evolve from focusing on isolated instances (cases) to understanding and addressing the broader networks (case clusters). Furthermore, we illustrate that this transition can be achieved with relative ease using currently available open-source tools.

\section{Background}
\label{sec:background}

We begin our discussion by elaborating the prevalent role of cryptoassets in cybercrime, placing a special emphasis on the relatively underexplored area of cybertrading fraud. This area, though significant in daily operations, has seen minimal coverage in existing scientific literature. Following this, we will briefly summarize known cryptoasset tracing techniques.

\subsection{Cryptoassets in cybercrime}

Cryptoassets nowadays play a significant role in all areas of cybercrime. Their misuse and the ensuing damage, is well-documented in academic literature. Foley et al.~\cite{Foley:2019a} discovered an illicit annual turnover of 76 billion USD. Anderson et al.~\cite{Anderson:2019a} estimated the direct costs of cybercrime involving cryptoassets to be around 2 billion USD, accounting for exchange hacks and direct crimes against individuals. According to Zhou et al.~\cite{Zhou:2022a}, between April 2018 and April 2022, users, liquidity providers, speculators, and operators of Decentralized Finance (DeFi) protocols collectively incurred losses totaling at least 3.24 billion USD. Oosthoek et al.~\cite{Oosthoek:2023a} reported that the revenue generated by Dark Web shops amounted to a minimum of 113 million USD, with sexual abuse (94 million USD) constituting the leading illicit category. Lastly, a recent industry report by Chainalysis~\cite{Chainalysis:2023a} estimated that illegal cryptoasset addresses received approximately 23 billion USD.

However, empirical evidence on the prevalence of cryptoassets in the prosecution of cybercrime is largely missing. To the best of our knowledge, only the US Securities and Exchange Commission provides some information and lists more than 100 enforcement actions against businesses involved in cryptoassets, and 16 trading suspensions related to cryptoassets\footnote{\url{https://www.sec.gov/spotlight/cybersecurity-enforcement-actions}}. 

\subsection{Cybertrading fraud}

Cybertrading fraud is a global issue involving misleading offers on trading platforms. Advertisements lure customers into investing significant sums in various financial instruments, including cryptocurrencies. However, the funds deposited are never genuinely invested; the visible trading platform and customer accounts are fabricated, almost always leading to total loss of the invested capital. The platforms operate with fraudulent intent, exploiting digitalization to hide their identities and locations. Investigations at the ZCB reveal that perpetrators use a global money laundering network to obscure the money flow\footnote{\url{https://www.justiz.bayern.de/gerichte-und-behoerden/generalstaatsanwaltschaft/bamberg/presse/2022/15.php}}. Victim-transferred funds are moved across at least two further levels to accounts of pseudo-companies at various banks or financial service providers worldwide.

Over the years, cybertrading fraud has become the most common offense linked to illicit cryptocurrency use in cases filed at the ZCB. Historically, victims of cybertrading fraud would transfer their deposit amounts to domestic or foreign bank accounts. The conversion of these funds from fiat currency to cryptocurrency, aimed at obscuring the payment trail, was typically executed by the perpetrators at a later stage. Consequently, transaction investigations were feasible at least at the initial levels through bank information. However, recent investigations reveal an increasing trend wherein victims' deposits are made directly using cryptocurrencies.

This shift can be attributed to the growing popularity and acceptance of cryptocurrencies among the general public. This trend is further encouraged by commission structures within call centers, as gleaned from broker interrogations. Brokers who convince victims to deposit using cryptocurrencies, due to the inherent difficulty in tracing the recipient or sender of such transactions, receive the highest commissions. Consequently, acquiring bank information is no longer a viable starting point for investigations. Instead, the analysis of cryptocurrency transactions has become crucial for initiating investigative leads.

\subsection{Connections between cases}

The intensive investigations of the ZCB in the past have shown that the globally active criminal groups usually operate a large number of platforms with different domains, often from various call centres in different countries. As soon as negative reviews from victims about a domain accumulate on the internet, the perpetrators offer their services via a new domain. Here it is important to recognise the connection between different platforms. This is often difficult because possible connections cannot be recognised at first sight.

Currently, however, almost all criminal police units in Bavaria are entrusted with the police investigations of the trading platform cases pending at the ZCB. After all, several criminal investigation departments are working on their respective individual proceedings in isolation due to their jurisdiction of residence for the reporting injured party. This dispersion at the police level, however, makes it extremely difficult to recognise platform connections and money laundering networks, especially at an early stage of the investigation.

Only an overall view of various findings from different proceedings currently allows factors such as matching recipient wallets or a jointly used money laundering network to be identified that speak for a connection between platforms. Therefore, identifying connections between cybertrading platforms as early as possible is of central importance in order to avoid duplicate investigations and to use the available resources efficiently.

\subsection{Following the money}

In recent years, the practice of tracing payment flows has gained significant traction as a forensic technique for investigating cryptoasset-related crimes. Numerous companies have emerged, offering tools capable of reconstructing payment flows using publicly available blockchain data and attribution tags that link cryptoasset addresses to real-world individuals or entities. This field of cryptoasset tracking and tracing has evolved into a thriving industry, attracting substantial investments from venture capital firms. Major tool providers, such as Chainalysis, TRM Labs, Elliptic, CipherTrace, and Merkle Science, have collectively received hundreds of millions of dollars in venture capital funding\footnote{According to data from Crunchbase, Chainalysis leads the pack with \$536.6 million, followed by TRM Labs with \$149.9 million, Elliptic with \$100 million, CipherTrace with \$45.1 million, and Merkle Science with \$25.6 million.}. 

While certain tools offer the capability to associate Bitcoin addresses with case information, this feature is typically limited to a select number of license holders within an organization. This limitation goes against the objective of identifying case connections at the earliest stage possible, as it necessitates broader availability of tools for criminal investigators within an institution. Furthermore, it is worth noting that law enforcement agencies are actively exploring alternatives to address this issue.  However, these databases are typically separate from forensic tools and require extensive manual reconciliation work, which can decrease motivation and overall adoption among investigators.

\section{Data and Methods}
\label{sec:data_methods}

This study aims to explore the connections between cases related to cybertrading fraud and sextortion. However, before proceeding with the systematic collection of the necessary data points to test our hypothesis, it is crucial to gain a deeper understanding of the patterns and terminology involved in money collection and laundering within these contexts.

\subsection{Conceptualizing money flows}

\begin{figure*}
     \centering
     \includegraphics[width=\textwidth]{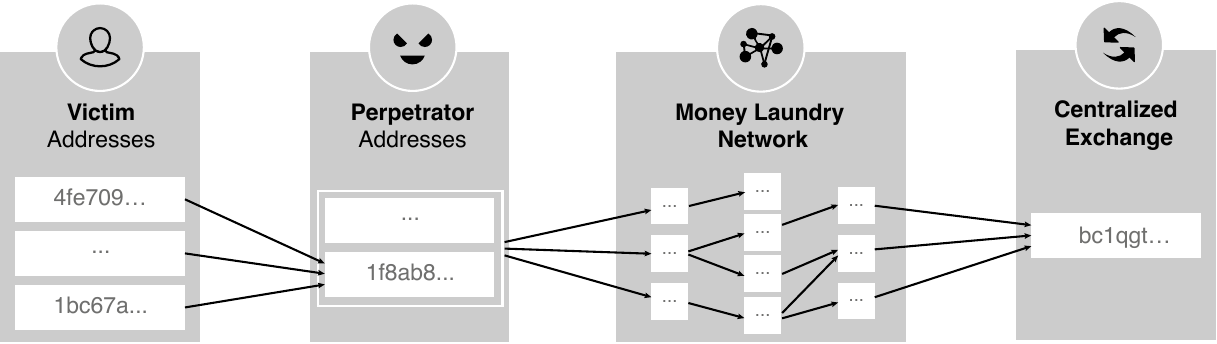}
     \caption{\textbf{Conceptual money flow}. Funds are transferred from victim to perpetrator addresses. From there, funds are funneled through a complex money laundry network before being reintegrated and cashed out at some centralized cryptoasset exchange.}
     \label{fig:conceptual_framework}
\end{figure*}

In the realm of cybertrading fraud, victims are commonly instructed to open accounts at well-established custodial cryptoasset exchanges. They transfer fiat currencies such as EUR or USD into these accounts and convert them into cryptoassets, typically Bitcoin. As illustrated in Figure~\ref{fig:conceptual_framework}, they subsequently transfer the cryptoassets from their accounts to one or more addresses controlled by the perpetrator. These on-chain transactions reveal a transfer from a ``victim address'' to a ``perpetrator'' address. The former is controlled by the victim, while the latter is under the control of the perpetrator. It is worth noting that the perpetrator often directs multiple victims to send payments to the same address, potentially transforming it into a ``collector address''.

Once the funds are collected, the perpetrator initiates the money laundering process, which typically involves various steps such as placement, layering, and integration. Cryptoasset exchanges, both centralized (CEXs) and decentralized (DEXs), are frequently utilized for placement and integration. Layering, on the other hand, often encompasses a multitude of addresses since their creation incurs negligible costs. In recent years, there has been a shift towards decentralized exchanges (DEXs) due to their lack of regulation and absence of Know-Your-Customer (KYC) measures (cf.~\cite{Auer:2023a}). Additionally, perpetrators, who are increasingly technologically adept, now transfer funds across multiple chains and automate their money laundering procedures to evade traceability. Consequently, the outcome is a complex network of money laundering that is arduous, if not impossible, to trace manually.

Much like cybertrading fraud, sextortion incidents can be viewed within a similar overarching framework. Victims likewise create accounts to transfer funds to the perpetrator, who then amasses these funds and starts the money laundering process.

\subsection{Case data collection}

Our data collection was based on filed cases of cybertrading fraud and sextortion at the Bavarian Central Office for the Prosecution of Cybercrime between \casePeriodStart and \casePeriodEnd. Each case is assigned a unique file number, such as BY1234-010123-22/6, which consists of the county code (BY for Bavaria), a distinctive identifier for the police station where the case was filed (1234), a sequential number within that station (010123), the year of filing (2022), and an integrity check number (6). In total, we gathered \fraudNrCases distinct cases of cybertrading fraud and \sextortionNrCases cases of sextortion.

For cybertrading fraud cases, we were able to obtain the victim addresses and the names of the cryptoasset exchanges where the victims opened their accounts. However, in our analysis of case connections, we focused on the addresses controlled by the perpetrators and disregarded victim addresses. In sextortion cases, we only have information about the cryptoasset addresses used in the emails received by the victims. Overall, we collected \fraudNrAddresses and \sextortionNrAddresses distinct addresses, all of which are Bitcoin addresses. To align with previous studies (see \cite{Paquet:2019b,Paquet:2019a,Oosthoek:2023a}), we refer to these addresses, which are controlled by the perpetrators, as \emph{seed addresses}.

\subsection{Address expansion and network abstraction}
\label{sec:clustering_network}

In the next phase, we expand the range of these addresses by computing address clusters, which encompass additional addresses likely controlled by the same real-world entity. To achieve this, we leverage the multi-input heuristics, pre-filtering CoinJoin transactions, using the toolset from GraphSense~\cite{Haslhofer:2021a}. We opt against other clustering heuristics due to their propensity for producing false positives~\cite{Moeser:2022a}. Through this augmentation process, we pinpoint 69 clusters containing 1563 addresses linked to cybertrading fraud, and 147 clusters with 968 addresses tied to sextortion spam. We label the addresses within these clusters as \emph{expanded addresses}. We can confidently infer that even though they weren't in the initial dataset, they are likely under the control of the perpetrators.

Once we have identified the pertinent clusters and addresses, we proceed to map them onto the entity graph. In this graph, each node corresponds to an address cluster, and an edge connecting two nodes signifies the combined financial transactions between the addresses encompassed within the source and target clusters. This abstraction allows us to trace the combined money inflows and outflows for a given perpetrator wallet.

Table~\ref{tbl:dataset} summarizes our seed and expanded datasets by case category. In total, it comprises \totalnrCases{} different cases including \totalnrAddresses{} perpetrator addresses. For \totalnrAddressesNoActivity{} of the addresses in the data-set we did not find any on-chain activity, after filtering them we end up with \totalCasesWithOnchainActivity{} cases where we identified activity on the bitcoin chain, which can be used to further analyse and connect the cases.

\begin{table*}
     \centering
     \caption[Summary statistics by case category]{Summary statistics of by case category.}
     \begin{tabular*}{.7\textwidth}{@{\extracolsep{\fill}}lrrrrrrr}
  \toprule
     & \multicolumn{2}{c}{Raw}
     & \multicolumn{2}{c}{Filtered}
     & \multicolumn{3}{c}{Expanded} \\
     \cline{2-3} \cline{4-5} \cline{6-8}
     & \# Cases & \# Addr.
     & \# Cases & \# Addr.
     & \# Cases & \# Addr. & \# Entities \\
  \midrule
  \bf Cyberfraud &\fraudNrCases{}&\fraudNrAddresses{}& \fraudCasesWithOnchainActivity{}&\fraudAddressesWithOnchainActivity{}& \fraudCasesWithOnchainActivity{}&\nrExpandedFraudAddresses{}&\nrExpandedFraudEntities{}\\[.5em]
  \bf Sextortion Spam &\sextortionNrCases{}&\sextortionNrAddresses{} & \sextortionCasesWithOnchainActivity{} & \sextortionAddressesWithOnchainActivity{}&  \sextortionCasesWithOnchainActivity{}& \nrExpandedSextortionAddresses{}& \nrExpandedSextortionEntities{}\\
  \midrule  
  $\sum$&\totalnrCases{} & \totalnrAddresses{} & \totalCasesWithOnchainActivity{} & 358 & \totalCasesWithOnchainActivity{} & \nrExpandedAddresses{} & \nrExpandedAddresses{}\\
  \bottomrule
\end{tabular*}
     \label{tbl:dataset}
\end{table*}

\subsection{Basic money flow statistics}

\begin{figure}
     \centering
     \resizebox{1\columnwidth}{!}{%
        \input{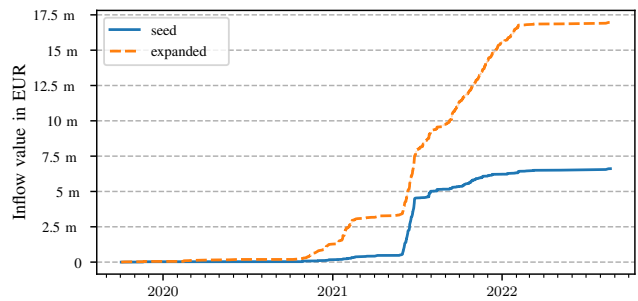}
     }
     \caption{\textbf{Cyberfraud inflows}. Cumulative payments received by addresses involved in cyberfraud cases.}
     \label{fig:inflow_fraud}
\end{figure}

\begin{figure}
     \centering
     \resizebox{1\columnwidth}{!}{%
        \input{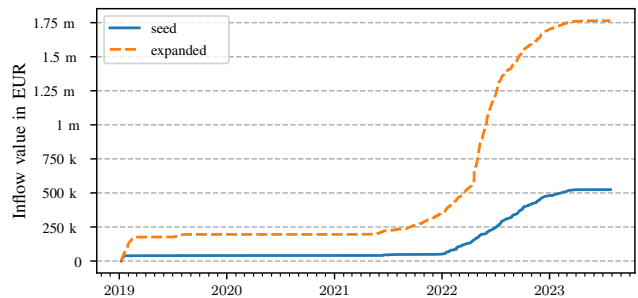}
     }
     \caption{\textbf{Sextortion spam inflows}. Cumulative payments received by addresses involved in sextortion spam cases.}
     \label{fig:inflow_sex}
\end{figure}

Having collected the pertinent addresses, we next extract the transactions involving these addresses from the publicly accessible Bitcoin blockchain. For the Cyberfraud cases, we found \nrtxfraudextended{} transactions from the period between \fraudcasePeriodStart{} and \fraudcasePeriodEnd{}. For the Sextortion spam cases, we pinpointed \nrtxsextortionextended{} transactions covering the period from \sexcasePeriodStart{} till \sexcasePeriodEnd{}. The inflow of funds for both the initial seed addresses and the expanded dataset are shown in Figures~\ref{fig:inflow_fraud} and \ref{fig:inflow_sex}. We removed transactions going directly to known cryptoasset exchanges and services entities.

By examining the payment values transferred to the perpetrator, we can glean insights into the pricing models of two distinct criminal schemes. Figure~\ref{fig:inflow_value_distribution_cyberfraud} illustrates the distribution of payments related to cyberfraud. It is evident that the majority of these payments fall within the low thousands, although there are some outliers exceeding 100k EUR. On the other hand, payments associated with sextortion spam, as depicted in Figure~\ref{fig:inflow_value_distribution_sextortion}, are considerably lower, typically under 1000 EUR.

\begin{figure}
     \centering
     \resizebox{1\columnwidth}{!}{%
        \input{figures/4_value_distribution_fraud.pgf}
     }
     \caption{\textbf{Cyberfraud payments}. Distribution of incoming cyberfraud transaction values.}
     \label{fig:inflow_value_distribution_cyberfraud}

\end{figure}

\begin{figure}
     \centering
     \resizebox{1\columnwidth}{!}{%
        \input{figures/4_value_distribution_sextortion.pgf}
     }
     \caption{\textbf{Sextortion spam payments}. Distribution of incoming sextortion spam transaction values.}
     \label{fig:inflow_value_distribution_sextortion}
\end{figure}

\section{Identifying case connections}
\label{sec:results}

\subsection{Common address heuristic}

The same cryptoasset addresses appearing across multiple cases are a robust and strong indication of linked activities. In our filtered dataset, which encompasses \totalCasesWithOnchainActivity{} cases, we observed that several cases already share identical addresses. By linking these cases, we identified \totalNrCaseClustersAddress{} distinct \emph{case clusters}, all of which have the same perpetrator address. Notably, the largest cluster identified using this straightforward address-matching approach connects \totalLargestCaseClusterAddress{} cases.

When considering case categories, it becomes evident that this method is particularly effective in linking sextortion spam cases collapsing \sextortionCasesWithOnchainActivity{} cases into \sextortionNrCaseClustersAddress{} case clusters. This outcome aligns with the known fact that sextortion spam campaigns are extensively automated, target victims globally, and often reuse cryptoasset addresses within and across campaigns~\cite{Paquet:2019b}. In contrast, for cybertrading fraud, there aren't any readily apparent links between cases. This can be explained by the intrinsic characteristics of the crime: perpetrators target victims on an individual basis and usually generate unique addresses for each victim.

Figure~\ref{fig:sextortion-address-cluster-breakdown} offers a detailed breakdown of the case clusters we identified. On the x-axis, we rank and enumerate case clusters by their size, while the y-axis illustrates the inflow for each cluster to highlight its significance. For example, the first column on the x-axis represents 84 case clusters, each of size one, indicating no established connections to other cases. The subsequent column represents 32 case clusters where two cases are connected by a shared address, and so forth. It's evident that only 84 cases (represented by the first column on the x-axis) remain isolated. For all the remaining cases, there's at least one identifiable link to another case, specifically a shared address. This suggests that even such a rudimentary approach enables us to associate roughly 93\% of the cases.

\begin{figure}
    \resizebox{1\columnwidth}{!}{%
        \input{figures/6_sextortion_address_case_cluster_distribution.pgf}
    }
    \caption{\textbf{Breakdown of sextortion spam case clusters based on common addresses}. The x-axis categorizes and enumerates cases by cluster size, while the y-axis represents the inflow for each cluster. Clusters with addresses managed by an exchange are marked by orange dots. For these, inflow values may represent hot wallets shared among multiple users and should be interpreted with caution.}
    \label{fig:sextortion-address-cluster-breakdown}
\end{figure}

\subsection{Common entity heuristic}

Clustering cases based on common addresses provides an intuitive and effective approach. However, this method does not account for the possibility that a single entity, such as a perpetrator, might control multiple cryptoasset addresses. As detailed in Section~\ref{sec:clustering_network}, we can group addresses that are likely controlled by the same real-world entity. We term such a group of addresses as an \emph{entity}. Extending the previous logic, we can elevate our analysis from an address-centric perspective to an entity-centric one, thereby linking cases based on these shared entities.

Applying this heuristic to our refined dataset reveals that \totalCasesWithOnchainActivity{} cases can be grouped into just \totalNrCaseClustersEntity{} case clusters. The most extensive of these clusters connects a remarkable \totalLargestCaseClusterEntity{} distinct cases. Much like the address-level analysis, the majority of connections are found within sextortion cases, consolidating them from \sextortionNrCaseClustersAddress{} down to \sextortionNrCaseClustersEntity{} case clusters. For fraud-related cases, our method resulted in the amalgamation of only two cases, leaving us with \fraudNrCaseClustersEntity{} clusters at the entity level. Notably, we didn't identify any overlaps between fraud and sextortion cases, a finding that aligns with the unique modus operandi characterizing each type of crime.

\begin{figure}
    \resizebox{1\columnwidth}{!}{%
        \input{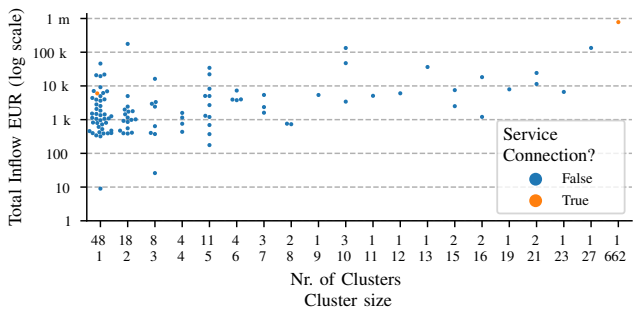}
    }
    \caption{\textbf{Breakdown of sextortion spam case clusters based on common entities}. The x-axis categorizes and enumerates cases by cluster size, while the y-axis represents the inflow for each cluster. Clusters with entities managed by an exchange are marked by orange dots.}
    \label{fig:sextortion-entity-cluster-breakdown}
\end{figure}

When we revisit the breakdown of case clusters for sextortion as depicted in Figure~\ref{fig:sextortion-entity-cluster-breakdown}, we observe that a mere 48 cases remain unlinked. This enhancement boosts our efficiency from the initial 92\% achieved with the basic common address heuristic to approximately 95\%.

Transitioning from addresses to entities through the multi-input heuristic proves generally effective~\cite{Harrigan:2016,Moeser:2022a}, provided specific preconditions are satisfied, such as the filtering of CoinJoin transactions. However, caution is advised when perpetrators deploy custodial addresses managed by service providers, like cryptoasset exchanges. Such actions could lead to the clustering of addresses belonging to unrelated entities. Therefore, we excluded entities tagged as exchange (11 instances) or having more than 10,000 addresses (2 instances). Consequently, we eliminated \clustersRemovedToBigOrExchange{} clusters from our inflow data, presuming that these addresses were associated with exchanges or other service providers in the ecosystem.

\subsection{Common collector heuristic}

Perpetrators frequently generate specific addresses designated for receiving funds from one or multiple victims. After collecting the funds, they often consolidate these victim payments into so-called \emph{collector addresses}. These funds are subsequently funneled through intricate money laundering networks. Notably, these collector addresses are directly connected to the perpetrator addresses, being just one hop away on the outgoing side. As a result, it is logical to infer that any address or entity transmitting funds to a collector address has some association with the criminal activity. Extending this logic to individual cases suggests that cases involving addresses or entities forwarding funds to the same collector address are likely interconnected or related

Applying the common collector heuristic on our sextortion spam cases, we observe a significant consolidation: our cases are distilled down to just \sextortionNrClustersWithCollectors{} case clusters. Only \sextortionNrSizeOneClusters{} cases remain isolated, lacking connections to other instances. This approach enables us to enhance the linkage of sextortion cases, elevating the connectivity rate from 95\% (as achieved using the common entity heuristic) to \emph{\pctLinkedCasesSextortion{}}.

When this method is employed for our cyberfraud cases, we observe notable enhancements in linkage. By focusing on the immediate entities that receive funds from recognized perpetrator addresses, our initial \fraudNrCaseClustersAddress{} cases condense down to \fraudNrClustersWithCollectors{} case clusters. While this approach may not be as potent as it was for the sextortion spam cases, it still successfully connects \emph{\pctLinkedCasesFraud{}} of our cases, leaving only \fraudNrSizeOneClusters{} without any connections.

\begin{figure*}
     \centering
     \includegraphics[width=\textwidth]{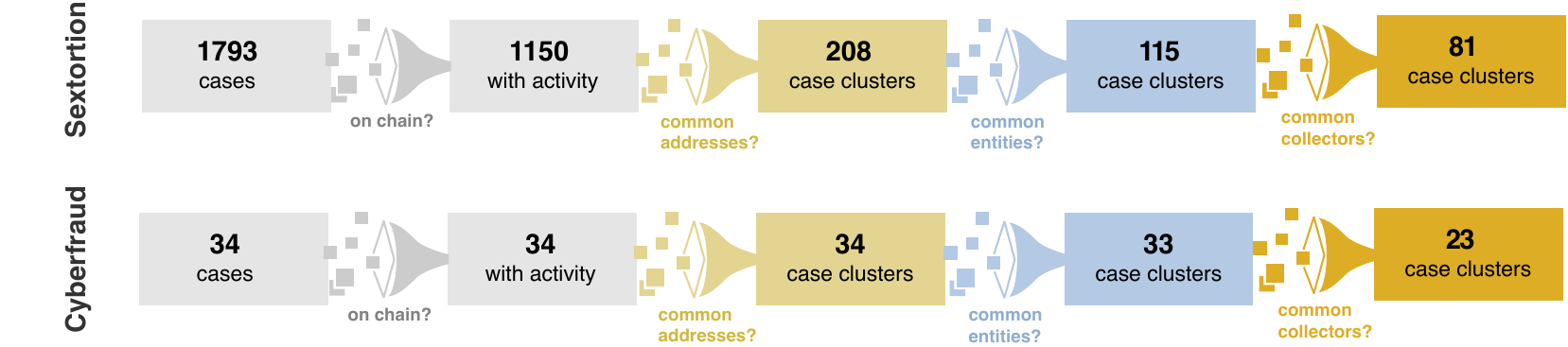}
     \caption{\textbf{Grouping cases into case clusters}. Collected sextortion spam and cyberfraud cases are first filtered for on-chain activity and then grouped into case clusters using the common address, common entity, and common collector heuristic.}
     \label{fig:case-connections-breakdown}
\end{figure*}

We must clearly proceed with caution when using collectors as a basis for linking cases. These collectors might not always directly be controlled by the perpetrators and sometimes be the targets of benign payments. Nevertheless, when treated with the necessary caution, collector addresses and entities can offer valuable insights into the interconnections between cases. Figure~\ref{fig:case-connections-breakdown} illustrates to what extent sextortion spam and cyberfraud cases can be grouped into smaller heuristics using the heuristics described above.

\subsection{Case networks}

\definecolor{casecolor}{HTML}{ccffcc}
\definecolor{addrcolor}{HTML}{ffe6cc}
\definecolor{clustercolor}{HTML}{ccccff}
\definecolor{collectorcolor}{HTML}{ffcccc}

\newcommand{\addressMarker}{\raisebox{0.5pt}{\tikz{\node[draw,scale=0.4,circle,fill=addrcolor!50!addrcolor](){};}}}
\newcommand{\caseMarker}{\raisebox{0.1pt}{\tikz{\node[draw,scale=0.8,circle,fill=casecolor!50!casecolor](){};}}}
\newcommand{\entityMarker}{\raisebox{0.5pt}{\tikz{\node[draw,scale=0.6,circle,fill=clustercolor!50!clustercolor](){};}}}
\newcommand{\collectorMarker}{\raisebox{0.5pt}{\tikz{\node[draw,scale=0.7,circle,fill=collectorcolor!50!collectorcolor](){};}}}

\begin{figure*}
     \centering\includegraphics[width=.7\textwidth]{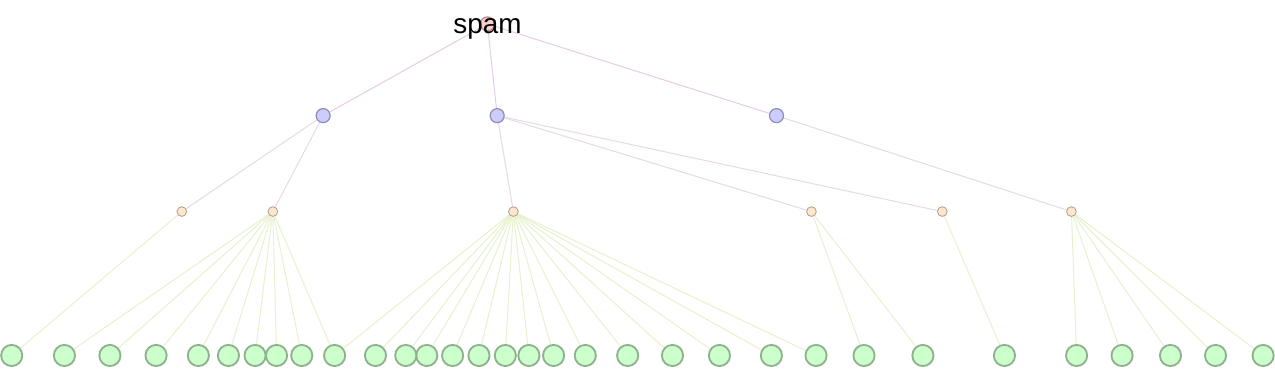}
    \caption{\textbf{Sextortion fraud: formation of the second largest cluster using the common collector heuristic}. The cluster shown connects 32 different cases. The color coding is as follows: green nodes \protect\caseMarker{} are cases, orange nodes are addresses \protect\addressMarker{}, purple nodes are entities \protect\entityMarker{} computed by the common entity heuristic and red nodes \protect\collectorMarker{} represent common collector entities.}
    \label{fig:fraud_network_with_hopconnection_second_largest_component}
\end{figure*}

\begin{figure}
    \centering
    \includegraphics[width=\columnwidth]{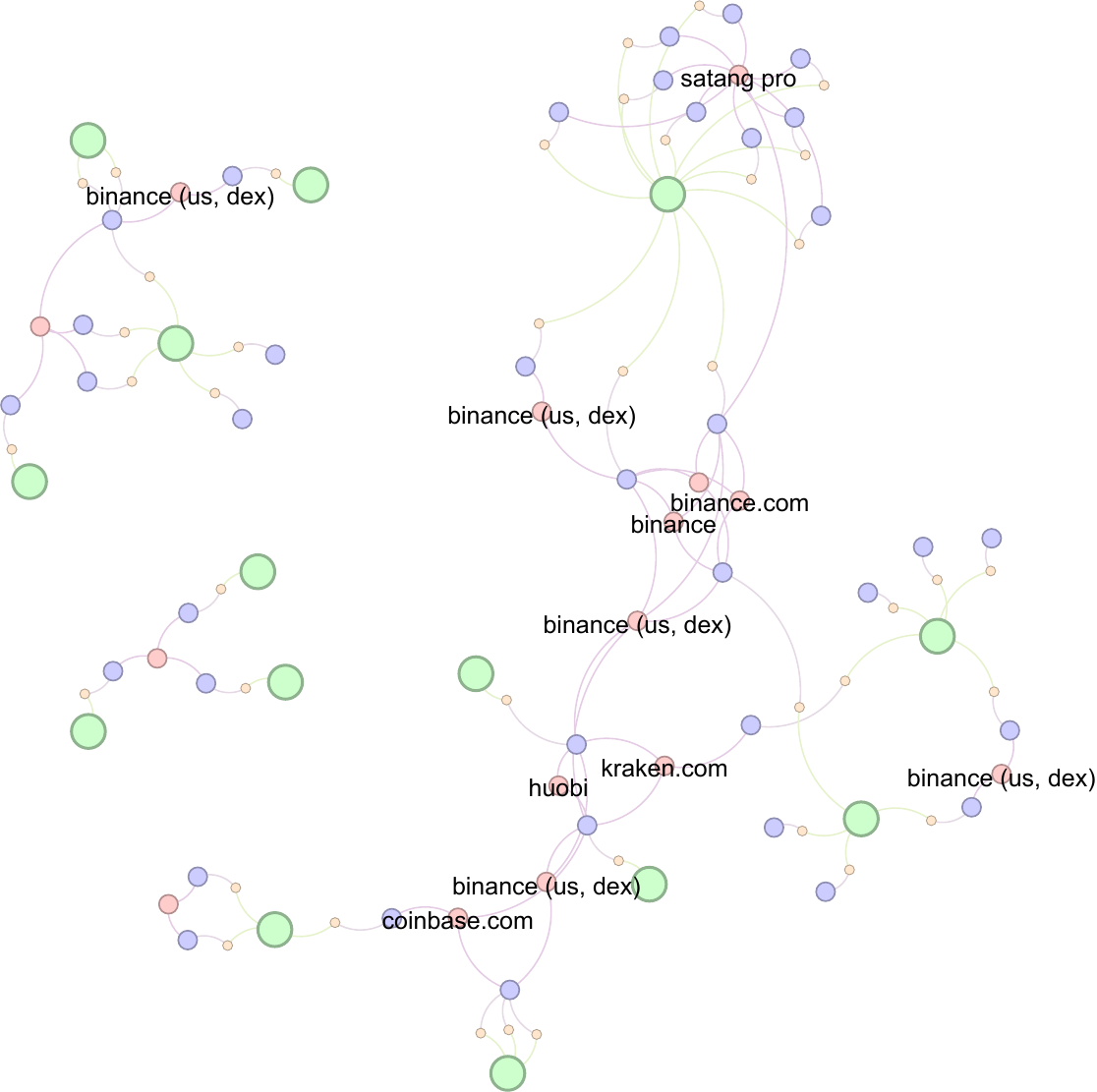}
    \caption{\textbf{Cybertrading fraud case network}. The color coding is as follows: green nodes \protect\caseMarker{} are cases, orange nodes are addresses \protect\addressMarker{}, purple nodes are entities \protect\entityMarker{} computed by the common entity heuristic and red nodes \protect\collectorMarker{} represent common collector entities.}
    \label{fig:fraud_network_with_hopconnection}
\end{figure}

\begin{figure}
    \centering
    \includegraphics[width=\columnwidth]{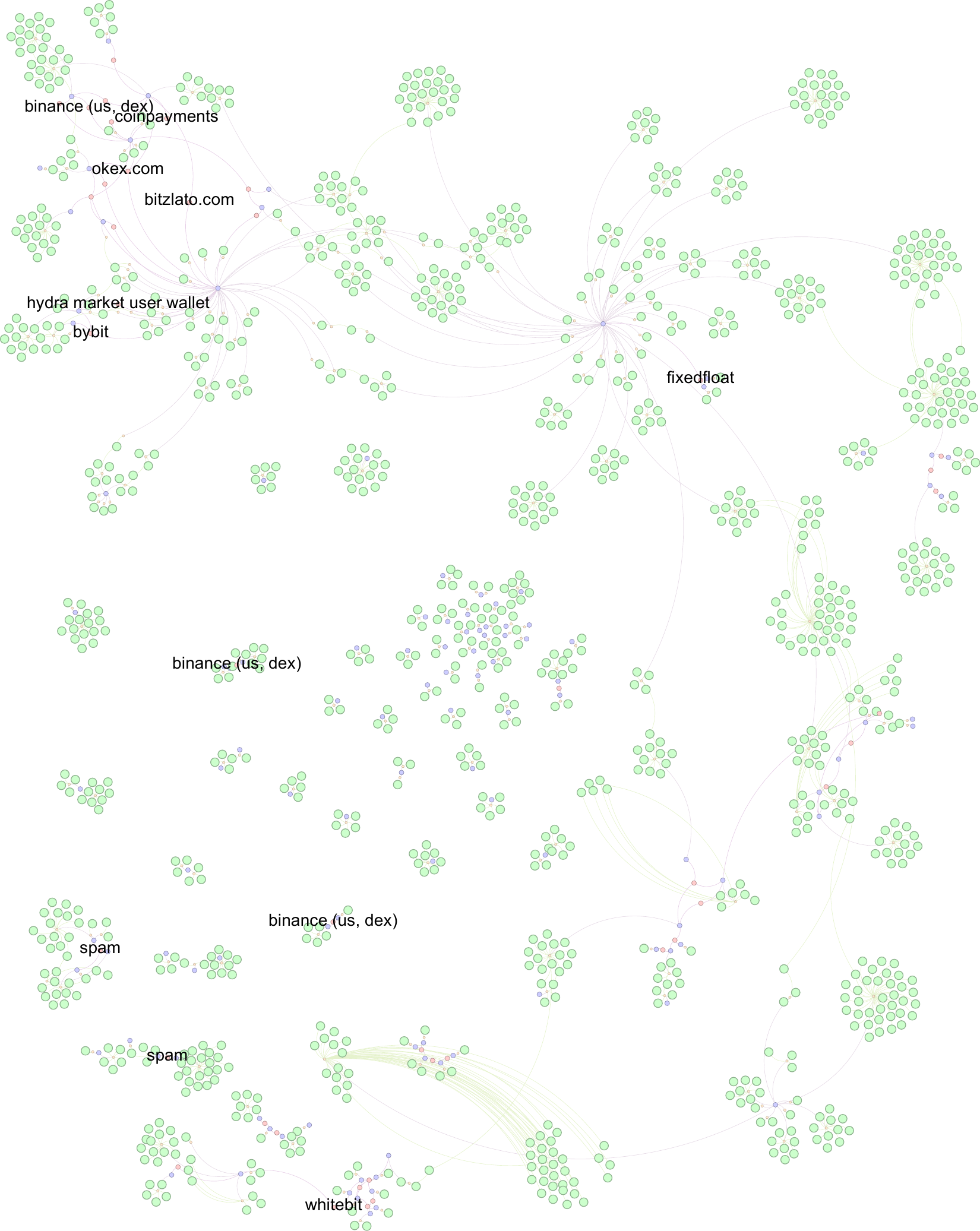}
    \caption{\textbf{Sextortion spam case network}. The color coding is as follows: green nodes \protect\caseMarker{} are cases, orange nodes are addresses \protect\addressMarker{}, purple nodes are entities \protect\entityMarker{} computed by the common entity heuristic and red nodes \protect\collectorMarker{} represent common collector entities.}
    \label{fig:sextortion_network_with_hopconnection}
\end{figure}

Figure \ref{fig:sextortion_network_with_hopconnection} and \ref{fig:fraud_network_with_hopconnection} visualize the complete case networks. Although, the two figures give a good overview of the connectedness in general it is hard to parse exactly how clusters are formed. Therefore, Figure \ref{fig:fraud_network_with_hopconnection_second_largest_component} illustrates the formation of one single cluster which is the the second largest case cluster in the sextortion cases. Its easy to see that one common collector (which is marked as being involved in a spam campaign) connects 3 smaller clusters formed via the common entity heuristic, and 5 clusters that are apparent from the address linking alone. In total the cluster links 32 different cases. 

\subsection{Overall damage estimation}

In Figures~\ref{fig:inflow_sex} and \ref{fig:inflow_fraud}, we presented the cumulative inflow into both the seed and expanded addresses for sextortion and cyberfraud cases, respectively. To prevent overestimation of victim flows, we excluded transactions that directly reached addresses likely associated with a service. As a result, the depicted values may be an underestimate of the actual flows.

In both crime categories, expanding our seed dataset more than doubles the observable inflow. The inflow from sextortion spam cases in the expanded dataset is estimated at approximately \num{1.7} million EUR, while the inflow from cyberfraud cases totals nearly \num{17} million EUR.

To estimate the number of potential victims, we can count the unique addresses found in incoming transactions. By focusing solely on the seed addresses, we identified at least \nrvictimsaddressessextortion{} unique addresses that sent money to a perpetrator in sextortion spam cases. Likewise, for fraud cases, we found \nrvictimsaddressesfraud{} unique sending addresses. However, this is a rudimentary estimate of the actual victim count. Some addresses might have executed multiple transactions to the perpetrator, which is often observed when funds are directly transferred from an exchange wallet.

\section{CaseConnect Tool}
\label{sec:caseconnect}

Our empirical research reveals that investigators often concurrently work on identical cases, unbeknownst to each other. The pressing question then arises: how can we facilitate these investigators in collaboratively pinpointing connections among these cases and subsequently disseminating this crucial information both intra- and inter-organizationally? The ultimate objective of implementing such a mechanism is to substantially enhance the efficiency of cryptoasset investigations. This can be achieved by detecting case interconnections at an early stage, ideally when the cases are first registered by police.

In the conception of our cryptoasset case management tool, we considered the following design guidelines:

\begin{enumerate}

	\item Recognizing that cases are not centrally filed but rather in a decentralized fashion by victims at individual police stations, our tool must be broadly accessible instead of being restricted to a handful of centralized units.

	\item The process of linking cryptoasset addresses to specific cases needs to be seamlessly incorporated within a foundational cryptoasset forensic workflow. This integration enables investigators to rapidly evaluate a case and determine potential links at the outset. The immediacy of this function is crucial, particularly when it comes to time-sensitive tasks like freezing assets.

	\item To preserve the integrity of the evidence in investigations, it is essential to avoid black box solutions. Additionally, adherence to the GDPR data minimization principle\footnote{\url{http://data.europa.eu/eli/reg/2018/1725/oj}} is imperative.

\end{enumerate}

In alignment with our overarching goal and design considerations, we opted to base our case management tool on the open-source cryptoasset analytics platform, GraphSense~\cite{Haslhofer:2021a}. Technically, our tool leverages the platform's customization mechanism and is implemented as add-on on top the foundational platform. This approach ensures its seamless integration within the comprehensive cryptoasset forensic workflow.

\begin{figure}
     \centering
     \includegraphics{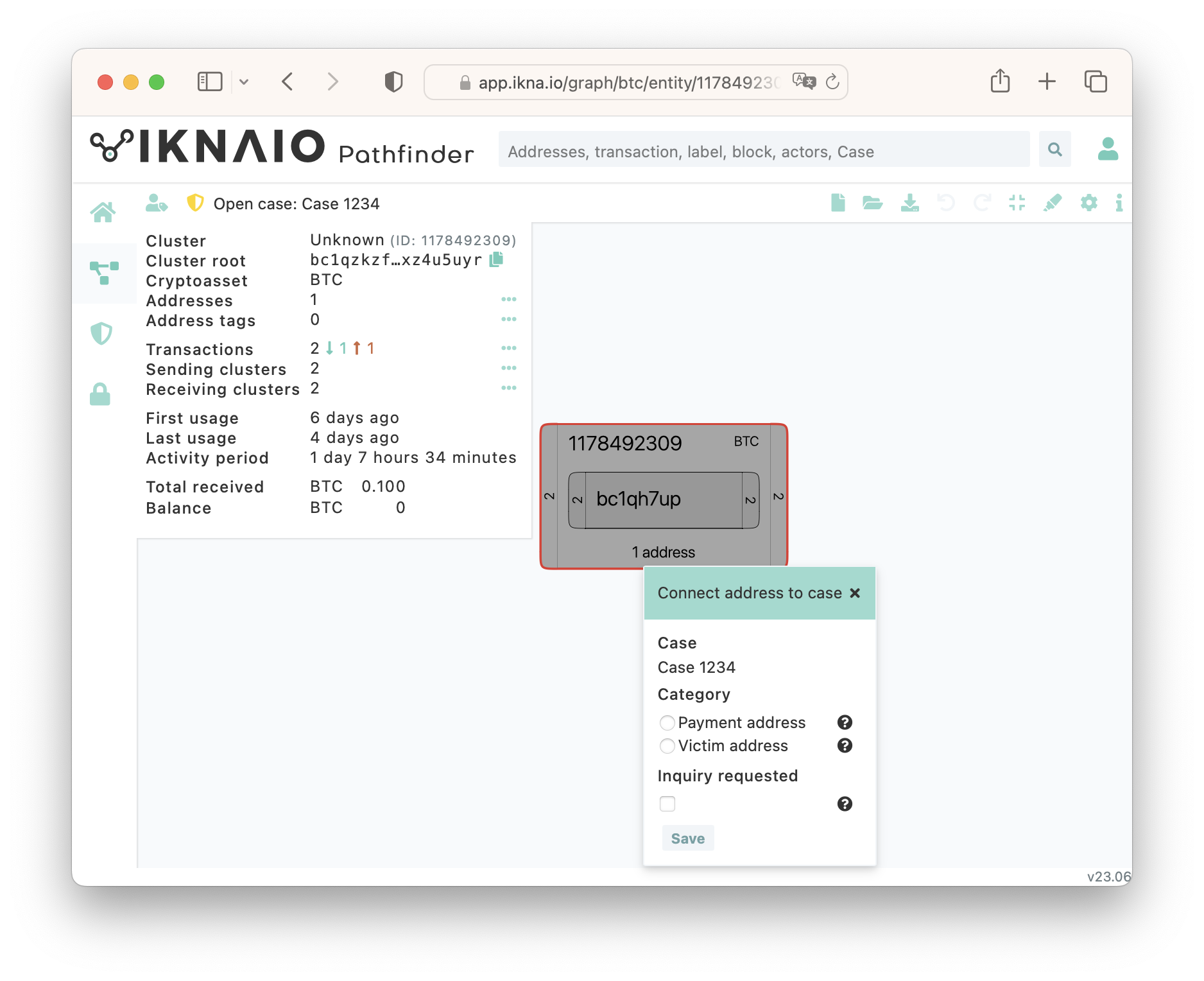}
     \caption{\textbf{Associating address with cases}. Users can create cases and associate addresses with these cases using a simple annotation mechanism.}
     \label{fig:case-annotation}
\end{figure}

Adhering to the principle of data minimization, our tool meticulously records only those data points pivotal to case linkage. Specifically, we capture case identifiers and map their associations with relevant cryptoasset addresses. Figure \ref{fig:case-annotation} provides an illustrative snapshot, depicting how users can associate addresses with respective cases through a straightforward annotation mechanism. To ensure clarity and eliminate ambiguities, we restrict annotations to only victim and perpetrator addresses, as these details are the primary facts presented when a case is filed. The detailed investigative trails and the resulting tracing graphs are archived within the case management system, accessible to other members within the same zone. Recognizing that investigations often necessitate information solicitations from law enforcement to cryptoasset exchanges, our system diligently notes which addresses have already been queried. This step is crucial to prevent redundant and time-consuming repeated requests.

\begin{figure}
     \centering
     \includegraphics{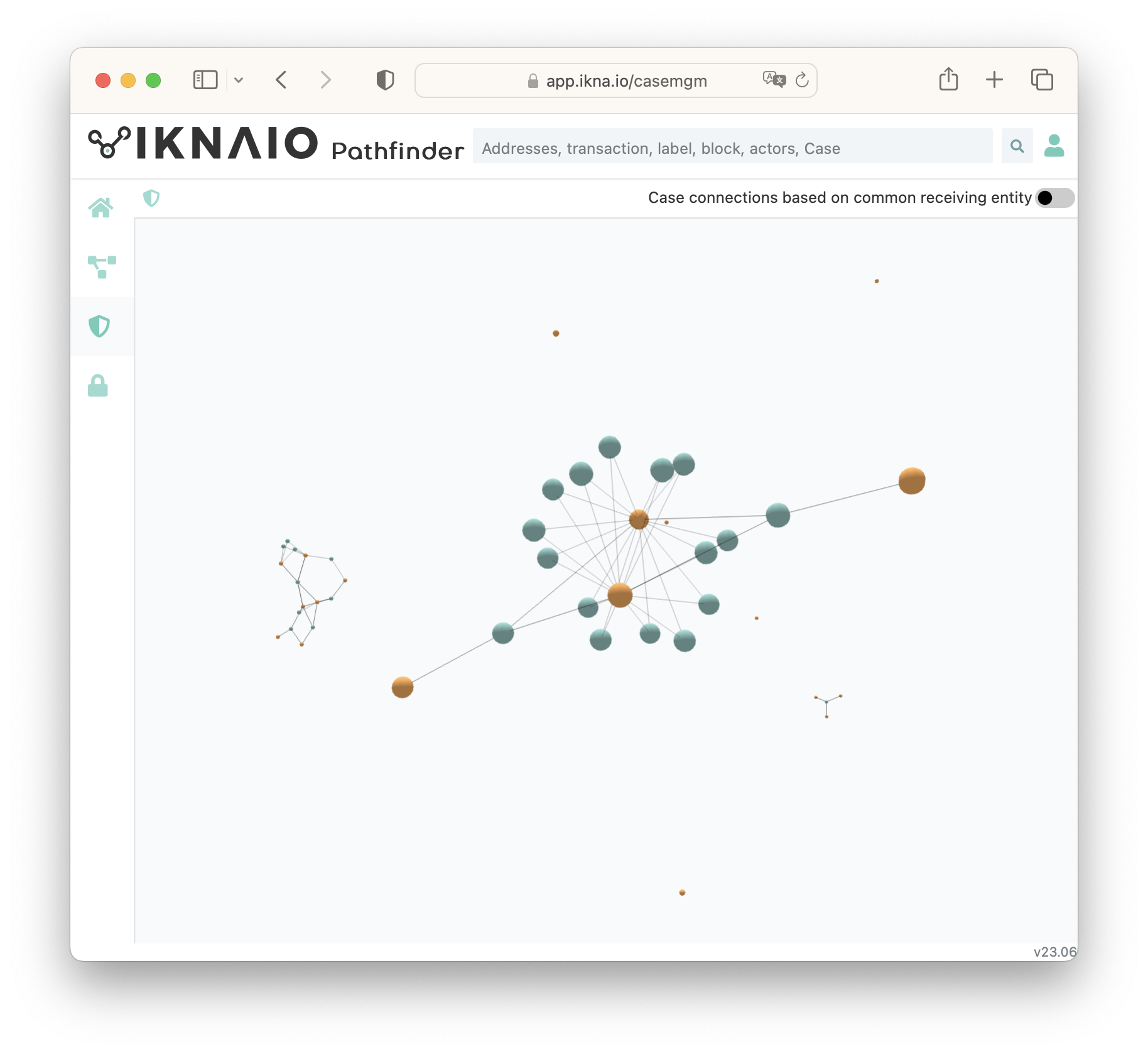}
     \caption{\textbf{Cases and case clusters}. Users can view all cases filed within a zone (organization) and visually inspect connections to other cases.}
     \label{fig:case-connections}
\end{figure}

Further bolstering data security and organization, case-related information is stored in an external relational database, which is organized into distinct \emph{zones}. Each zone can be envisioned as an institution, where users have the privilege of viewing annotations related to cases within their specific zone. There's also flexibility built into the system, allowing for the potential readability across zones. Figure~\ref{fig:case-connections} offers a visual representation of the myriad of cases and demonstrates their aggregation into what we term as \emph{case clusters}.

\section{Discussion}\label{sec:discussion}


Our study is motivated by the pressing challenge faced by prosecutors in Bavaria, who are confronted with an escalating number of cryptoasset-related cases. In their daily work, they observed that investigators often unintentionally delve into overlapping or interconnected cases, leading us to postulate that identifying these links could substantially enhance investigation efficiency. To validate this, we analyzed sample cases from two distinct crime areas, namely cybertrading fraud and sextortion spam, aiming to establish heuristics for detecting inter-case connections. The findings from our study compellingly indicate the widespread nature of such connections, thus confirming our hypothesis. This is also in-line with previous findings showing that concentration appears to emerge frequently in cybercrime~\cite{clayton:2015}.

In response, we have developed a simple yet potent tool to unveil these case linkages. If adopted extensively, this tool can not only streamline investigative processes but also render economic benefits. Specifically, by focusing on clusters of interlinked cases instead of individual ones, we can achieve tangible cost savings at both the police and prosecutorial levels, allowing for a more targeted use of public funds.


While our research has provided insightful findings and a solution for a pressing challenge, it does come with certain acknowledged limitations: notably, the sample size, particularly for cybertrading fraud, is somewhat limited and geographically confined to Bavaria. Additionally, our methods using common entity and common collector heuristics operate under the assumption that perpetrators utilize non-custodial wallets to amass and collect payments from victims. To avoid false positives, we have implemented safeguards, such as filtering out recognized exchanges and excluding clusters with substantial address counts, as a practical approach to distinguish services like exchanges. Nevertheless, the possibility of a few false positives remains, which may marginally diminish the reported numbers.


We also see several potential avenues for future work: cryptoassets have unmistakably emerged as universal payment methods across various domains of cybercrime. Consequently, an immediate extension of our research should include other cybercrime arenas such as ransomware, child sexual abuse, and romance scams. Additionally, the scope of our study was limited to Bavaria, a mere fraction when considering that it is one among 16 provinces in Germany, which itself is just one European country, hinting at the vastness of the global stage. This microscopic focus underscores the vast potential we can unleash by expanding our methodology and approach across international boundaries and jurisdictions, especially given the pervasive and global implications of cybercrime.

On a broader scale, our study highlights a prevailing challenge: the super-linear growth of cyber-related incidents and the resulting investigations that arise in combating cyber and cyber-enabled crime. While augmenting human resources might seem like a solution, this approach realistically scales linearly at best and often less efficiently due to the organizational challenges that come with expanding teams. In the dynamic realm of cyber threats, the significance of automated assistance, combined with network-oriented thinking, addressing inherent complexities, and leveraging understood connections, will be of paramount importance.

\section{Conclusions}
\label{sec:conclusions}

Cybertrading fraud and sextortion spam are witnessing an alarming surge in case numbers. Given the limited resources at hand, it is important for law enforcement to reevaluate and innovate their approach to these challenges. Merely augmenting human resources and purchasing additional licenses for centralized units is not a sustainable solution, as such strategies often scale linearly at best. Our research clearly showcases the interconnectivity and overlap among these cases. We are confident that there is a potential for tremendous efficiency improvements when this approach is extrapolated across different crime domains and transcends international borders. This study underscores the necessity of rethinking entrenched procedures, embracing a network-centric perspective, and deploying supportive tools on a broader scale. These systems, designed to discern connections and foster knowledge-sharing, represent the future of efficient and collaborative investigations. In essence, by harnessing these insights, investigators can optimize their workflows and benefit from the synergistic effects of shared knowledge and expertise.

\section*{Acknowledgements}

The authors would also like to thank Lukas Knorr and Barbara Kr\"{u}ll for their support in gathering the data for this study. This work is partially funded by the Austrian Research Promotion Agency and the Austrian security research programme KIRAS of the Federal Ministry of Finance (BMF) under the project DeFiTrace (905300). 

\clearpage{}

\bibliographystyle{IEEEtran}
\bibliography{references}

\end{document}